\documentclass{jps-cp}
\usepackage{txfonts} 

\title{Effects of strong magnetic fields on neutron $^{3}P_{2}$ superfluidity with spin-orbit interactions}

\author{Shigehiro \textsc{Yasui},  Chandrasekhar \textsc{Chatterjee}, and Muneto \textsc{Nitta}}

\inst{Department of Physics \& Research and Education Center for Natural Sciences, Keio University, Hiyoshi 4-1-1, Yokohama, Kanagawa 223-8521, Japan}

\email{yasuis@keio.jp}

\recdate{October 5, 2016}

\abst{We discuss neutron $^{3}P_{2}$ phases in the core of neutron stars in strong magnetic field (magnetars). The neutron $^{3}P_{2}$ pairing provides a wide variety of condensates, such as the uniaxial nematic and (D$_{2}$ and D$_{4}$) biaxial nematic, with different symmetries stemming from the combinations of spin and momentum. Based on the spin-orbital angular momentum coupling and the spin-magnetic field coupling of the neutrons, we derive the Ginzburg-Landau equation containing higher order terms of the magnetic field. We investigate the phase diagram of the neutron $^{3}P_{2}$ superfluidity, and find that the D$_{2}$ biaxial nematic phase is extended by the higher order terms of the magnetic field. We also discuss the thermodynamic properties, the heat capacity and the spin susceptibility.}

\kword{p-wave superfluidity, neutron stars, Ginzburg-Landau equation}

\begin{document}
\maketitle

\section{Introduction}

Neutron stars are interesting astrophysical objects to study high density state of the strong interaction (see e.g. Ref.~\cite{Baym:2017whm}).
As for the inner structures in the neutron stars, it has been discussed that the neutron gas forms the $^{3}P_{2}$ superfluidity (see e.g. Refs.~\cite{Chamel:2017}).
It is known from the high-energy experiments that the attraction in the $^{3}P_{2}$ channel is provided by the $LS$ potential at the energy scales relevant to the densities higher than the normal nuclear matter~\cite{neutron_superfluidity_LS}.
In the $^{3}P_{2}$ channel, there is a wide variety of types of the condensates which cannot be seen in normal $S$-wave superfluidity: the nematic phase~\cite{neutron_superfluidity_3P2,Masuda:2015jka,Masuda:2016vak}, the cyclic phase and the ferromagnetic phase and so on.
The nematic phase is further classified to the uniaxial nematic (UN) phase with $\mathrm{U}(1)$ symmetry, and the biaxial nematic (BN) phase with D$_{2}$ or D$_{4}$ symmetry.
Interestingly, it has been shown that the neutron $^{3}P_{2}$ superfluidity has topological properties~\cite{Masuda:2015jka,Masuda:2016vak,Mizushima:2016fbn,Chatterjee:2016gpm}.
They provide various thermodynamic properties which should give an impact on the astrophysical observations of the neutron stars.

The purpose in the presentation is to investigate the phase diagram of the neutron $^{3}P_{2}$ superfluidity in the neutron stars with the strong magnetic field (magnetars).
The magnetars have very strong magnetic field around $10^{15}$ G (or $10^{11}$ T) at the surface,
 which is about hundred times larger than that in normal neutron stars.
The strong magnetic field can affect the neutron $^{3}P_{2}$ superfluidity, because the neutron can couple to the magnetic field through the spin-magnetic field interaction.
The fundamental equation of the neutron $^{3}P_{2}$ superfluidity is the Bogoliubov-de Gennes (BdG) equation (see~\cite{Mizushima:2016fbn} for a recent work).
In the literature, the GL equation has been often used as an effective theory of the neutron $^{3}P_{2}$ pairing around the transition region from the normal state to the superfluid state.
So far, however, the expansion about the magnetic field was limited only in the lowest order, and it was not clear how the GL equation is applicable in the strong magnetic field (see e.g. \cite{Masuda:2015jka}).
In the present study, we extend the previous GL equation to include the higher order terms of the magnetic field, and investigate the change of the phase diagram of the neutron $^{3}P_{2}$ superfluidity~\cite{Yasui:2018tcr}.

\section{Ginzburg-Landau equation}

Considering the microscopic view that two neutrons are interacting attractively in the $^{3}P_{2}$ channel by the $LS$ potential,
we give the Lagrangian of the neutron ($\varphi$) as
\begin{eqnarray}
  {\cal L}[\varphi]
=
   \varphi(t,\vec{x})^{\dag} \biggl( i\partial_{t} - \frac{\vec{\nabla}^{2}}{2m} - \mu + \vec{\mu}_{n} \!\cdot\! \vec{B} \biggr) \varphi(t,\vec{x})
+ G \sum_{a,b} T^{ab}(t,\vec{x})^{\dag}T^{ab}(t,\vec{x}),
\label{eq:action_tensor_0}
\end{eqnarray}
where $m=939$ MeV is a neutron mass and $\mu$ is the chemical potential of the neutron gas, and $- \vec{\mu}_{n} \!\cdot\! \vec{B}$ is the interaction term for the magnetic moment of a neutron $\vec{\mu}_{n}=\gamma_{n}\vec{\sigma}/2$ ($\gamma_{n}=1.2 \times 10^{-13}$ MeV/T the gyromagnetic ratio in natural units, $\hbar=c=1$) and the magnetic field $\vec{B}$~\cite{neutron_superfluidity_LS,Masuda:2015jka,Masuda:2016vak,Chatterjee:2016gpm,Yasui:2018tcr}.
The second term in the right-hand-side denotes the neutron-neutron interaction with the coupling constant $G>0$ (attraction), which is expressed by a symmetric and traceless tensor operator defined by
$
 T^{ab}(t,\vec{x})
= \frac{1}{2} \Bigl( \phi^{ab}(t,\vec{x}) + \phi^{ba}(t,\vec{x}) \Bigr) - \frac{1}{3} \delta^{ab} \sum_{c} \phi^{cc}(t,\vec{x})
$
($a,b=1,2,3$; spin and space directions). Here $\phi^{ab}(t,\vec{x})$ is the pairing function defined by
$
    \phi^{ab}(t,\vec{x})
= - \varphi(t,\vec{x})^{t} \Sigma^{a\dag} \bigl( \nabla^{b}_{x} \varphi(t,\vec{x}) \bigr)
$
with $\Sigma^{a} = i\sigma^{a}\sigma^{2}$ and $\nabla^{b}_{x}=\partial/\partial x^{b}$.
We adopt the bosonization technique for the neutron $^{3}P_{2}$ pairing by introducing the condensate $A^{ab}$
 as the mean-field $-G\bigl\langle T^{ab}(t,\vec{x}) \bigr\rangle$ for the tensor operator.
Applying the one-loop approximation and the quasi-classical approximation in the momentum integrals for the neutrons,
we obtain the free energy density~\cite{Yasui:2018tcr}:
\begin{eqnarray}
  f[{A}] = f_{0} + f_{6}^{(0)}[{A}] + f_{2}^{(\le4)}[{A}] + f_{4}^{(\le2)}[{A}] + {\cal O}(B^{m}{A}^{n})_{m+n\ge7}.
\label{eq:eff_pot_coefficient02_f}
\end{eqnarray}
Here $f_{0}$ is the term irrelevant to the condensate.
The following terms are:
\begin{eqnarray}
 f_{6}^{(0)}[{A}]
&=&
  K^{(0)}
  \Bigl(
        \nabla_{xi} {A}^{ba\ast}
        \nabla_{xi} {A}^{ab}
     + \nabla_{xi} {A}^{ia\ast}
        \nabla_{xj} {A}^{aj}
     + \nabla_{xi} {A}^{ja\ast}
        \nabla_{xj} {A}^{ai}
  \Bigr)
\nonumber \\ &&
+ \alpha^{(0)}
   \mathrm{tr}\bigl( {A}^{\ast} {A} \bigr)
+ \beta^{(0)}
   \Bigl(
        \mathrm{tr}\bigl( {A}^{\ast} {A} \bigr) \mathrm{tr}\bigl( {A}^{\ast} {A} \bigr)
      - \mathrm{tr}\bigl( {A}^{\ast} {A}^{\ast} {A} {A} \bigr)
   \Bigr)
\nonumber \\ &&
+ \gamma^{(0)}
   \Bigl(
         - 3 \, \mathrm{tr}\bigl( {A} {A}^{\ast} \bigr) \, \mathrm{tr}\bigl( {A} {A} \bigr) \, \mathrm{tr}\bigl( {A}^{\ast} {A}^{\ast} \bigr)
        + 4 \, \mathrm{tr}\bigl( {A} {A}^{\ast} \bigr) \, \mathrm{tr}\bigl( {A} {A}^{\ast} \bigr) \, \mathrm{tr}\bigl( {A} {A}^{\ast} \bigr)
              \nonumber \\ && \hspace{3em} 
        + 6 \, \mathrm{tr}\bigl( {A}^{\ast} {A} \bigr) \, \mathrm{tr}\bigl( {A}^{\ast} {A}^{\ast} {A} {A} \bigr)
      + 12 \, \mathrm{tr}\bigl( {A}^{\ast} {A} \bigr) \, \mathrm{tr}\bigl( {A}^{\ast} {A} {A}^{\ast} {A} \bigr)
              \nonumber \\ && \hspace{3em} 
         - 6 \, \mathrm{tr}\bigl( {A}^{\ast} {A}^{\ast} \bigr) \, \mathrm{tr}\bigl( {A}^{\ast} {A} {A} {A} \bigr)
         - 6 \, \mathrm{tr}\bigl( {A} {A} \bigr) \, \mathrm{tr}\bigl( {A}^{\ast} {A}^{\ast} {A}^{\ast} {A} \bigr)
              \nonumber \\ && \hspace{3em} 
       - 12 \, \mathrm{tr}\bigl( {A}^{\ast} {A}^{\ast} {A}^{\ast} {A} {A} {A} \bigr)
      + 12 \, \mathrm{tr} \bigl( {A}^{\ast} {A}^{\ast} {A} {A} {A}^{\ast} {A} \bigr)
        + 8 \, \mathrm{tr}\bigl( {A}^{\ast} {A} {A}^{\ast} {A} {A}^{\ast} {A} \bigr)
   \Bigr),
\label{eq:eff_pot_w0_coefficient02_f}
\\
   f_{2}^{(\le4)}[{A}]
&=&
      \beta^{(2)}
      \vec{B}^{t} {A} {A}^{\ast} \vec{B}
+ \beta^{(4)}
   |\vec{B}|^{2}
   \vec{B}^{t} {A} {A}^{\ast} \vec{B},
\label{eq:eff_pot_B4w2_coefficient02_f}
\\
   f_{4}^{(\le2)}[{A}]
&=&
  \gamma^{(2)}
  \Bigl(
       - 2 \, |\vec{B}|^{2} \, \mathrm{tr}\bigl( {A} {A} \bigr) \, \mathrm{tr}\bigl( {A}^{\ast} {A}^{\ast} \bigr)
      - 4 \, |\vec{B}|^{2} \, \mathrm{tr}\bigl( {A} {A}^{\ast} \bigr) \, \mathrm{tr}\bigl( {A} {A}^{\ast} \bigr)
      + 4 \, |\vec{B}|^{2} \, \mathrm{tr}\bigl( {A} {A}^{\ast} {A} {A}^{\ast} \bigr)
            \nonumber \\ && \hspace{2em}
      + 8 \, |\vec{B}|^{2} \, \mathrm{tr}\bigl( {A} {A} {A}^{\ast} {A}^{\ast} \bigr)
        + \vec{B}^{t} {A} {A} \vec{B} \, \mathrm{tr}\bigl( {A}^{\ast} {A}^{\ast} \bigr)
       - 8 \, \vec{B}^{t} {A} {A}^{\ast} \vec{B} \, \mathrm{tr}\bigl( {A} {A}^{\ast} \bigr)
         + \vec{B}^{t} {A}^{\ast} {A}^{\ast} \vec{B} \, \mathrm{tr}\bigl( {A} {A} \bigr)
            \nonumber \\ && \hspace{2em}
      + 2 \, \vec{B}^{t} {A} {A}^{\ast} {A}^{\ast} {A} \vec{B}
      + 2 \, \vec{B}^{t} {A}^{\ast} {A} {A} {A}^{\ast} \vec{B}
       - 8 \, \vec{B}^{t} {A} {A}^{\ast} {A} {A}^{\ast} \vec{B}
       - 8 \, \vec{B}^{t} {A} {A} {A}^{\ast} {A}^{\ast} \vec{B}
  \Bigr).
\label{eq:eff_pot_B2w4_coefficient02_f}
\end{eqnarray}
The concrete expressions of the coefficients ($K^{(0)}$, $\alpha^{(0)}$, $\dots$) are found in Ref.~\cite{Yasui:2018tcr}.
In the above equations, the terms with $\beta^{(4)}$ and $\gamma^{(2)}$ are the new terms stemming from the higher order of the magnetic field.
We use the parameter setting:
the critical temperature $T_{c0}=0.2$ MeV,
the nuclear matter density $n=0.17$ fm$^{-3}$ (the Fermi momentum $p_{F}=338$ MeV),
the Landau parameter $F_{0}^{a}=-0.75$~\cite{Masuda:2016vak,Chatterjee:2016gpm,Yasui:2018tcr}.

\section{Numerical results}

\begin{figure}[tb]
\begin{center}
\includegraphics[scale=0.5]{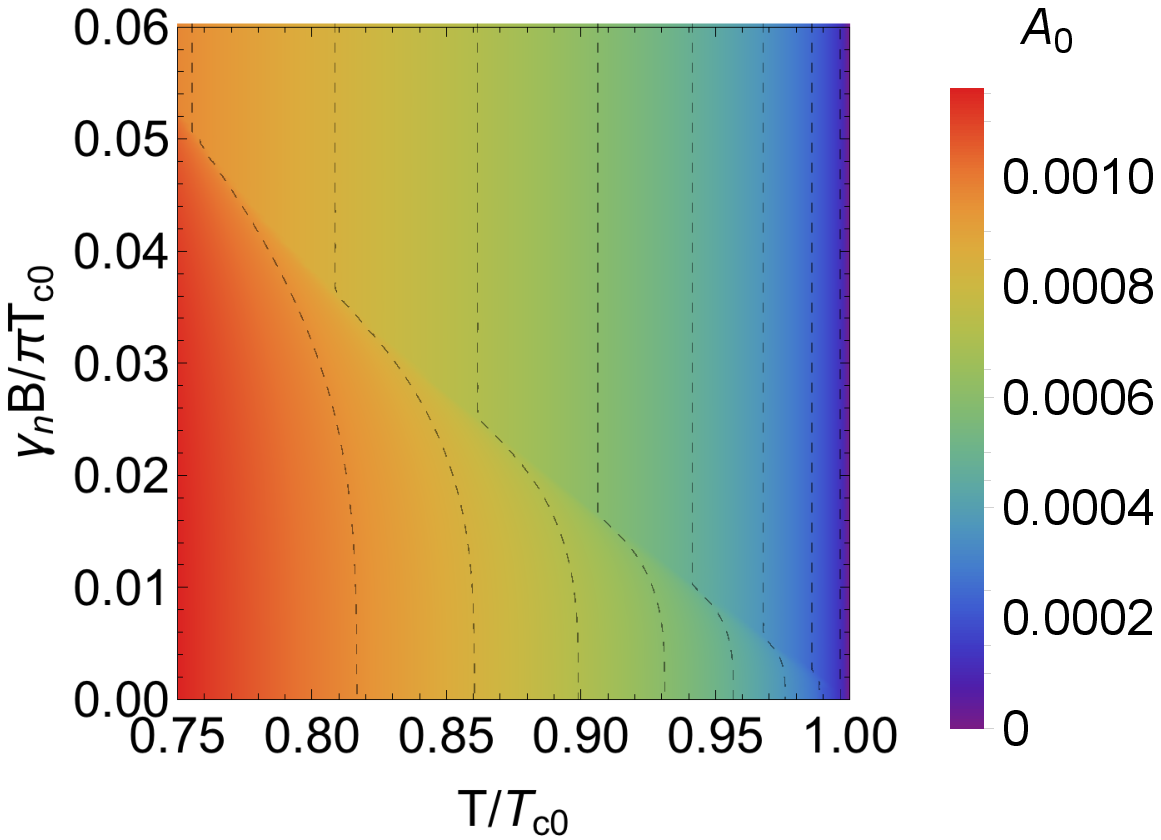}
\hspace{1em}
\includegraphics[scale=0.5]{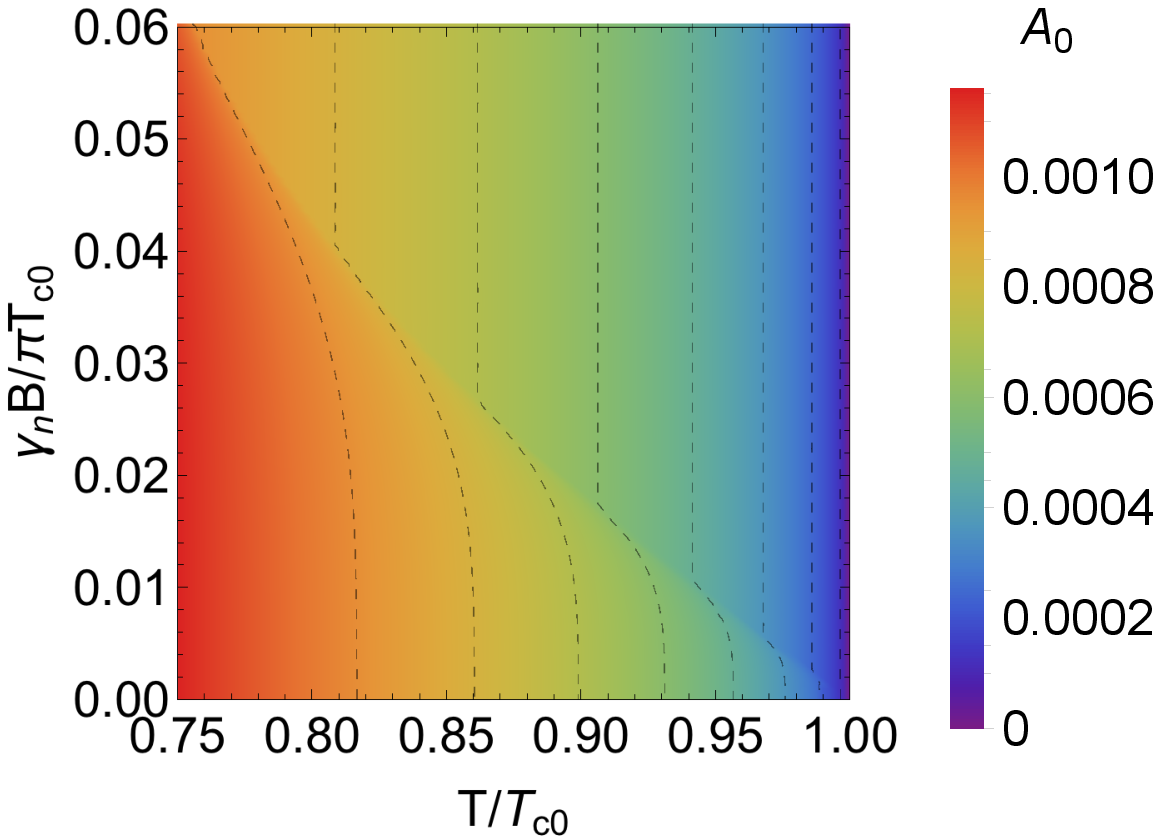}
\vspace{1em}
\includegraphics[scale=0.5]{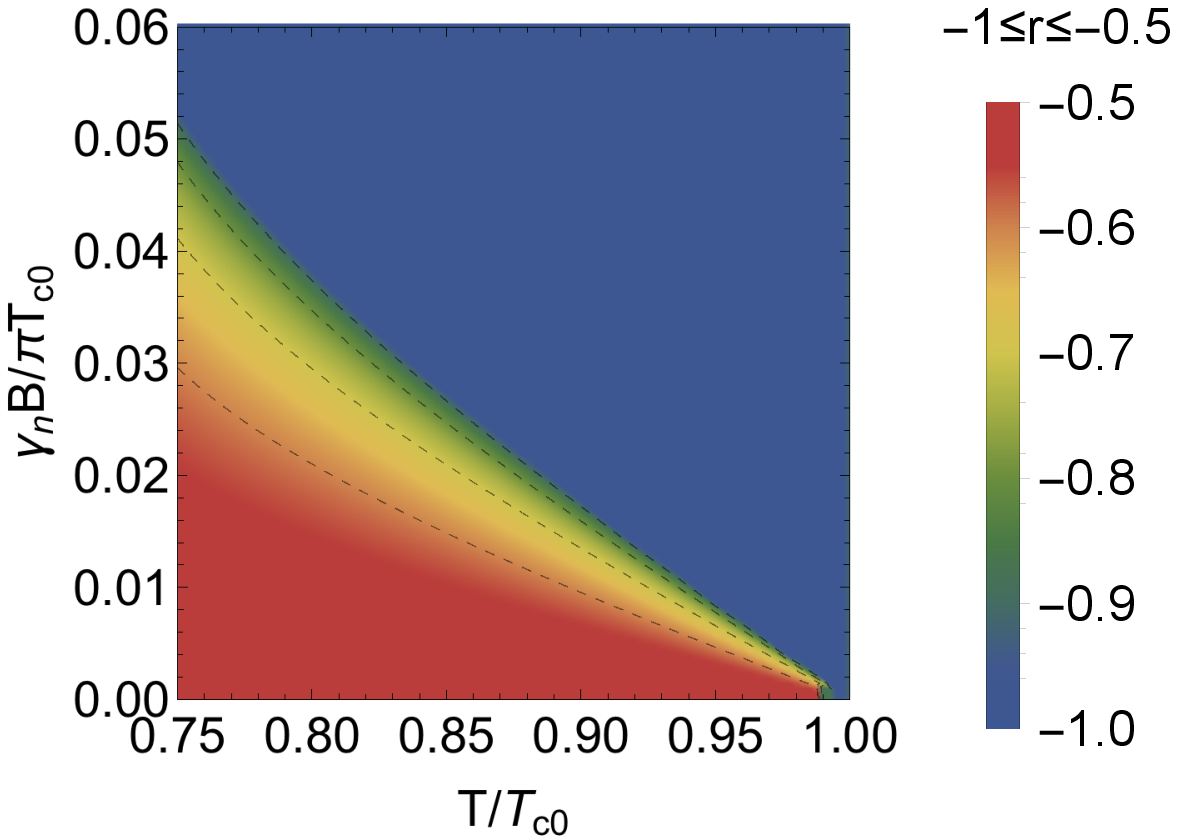}
\hspace{1em}
\includegraphics[scale=0.5]{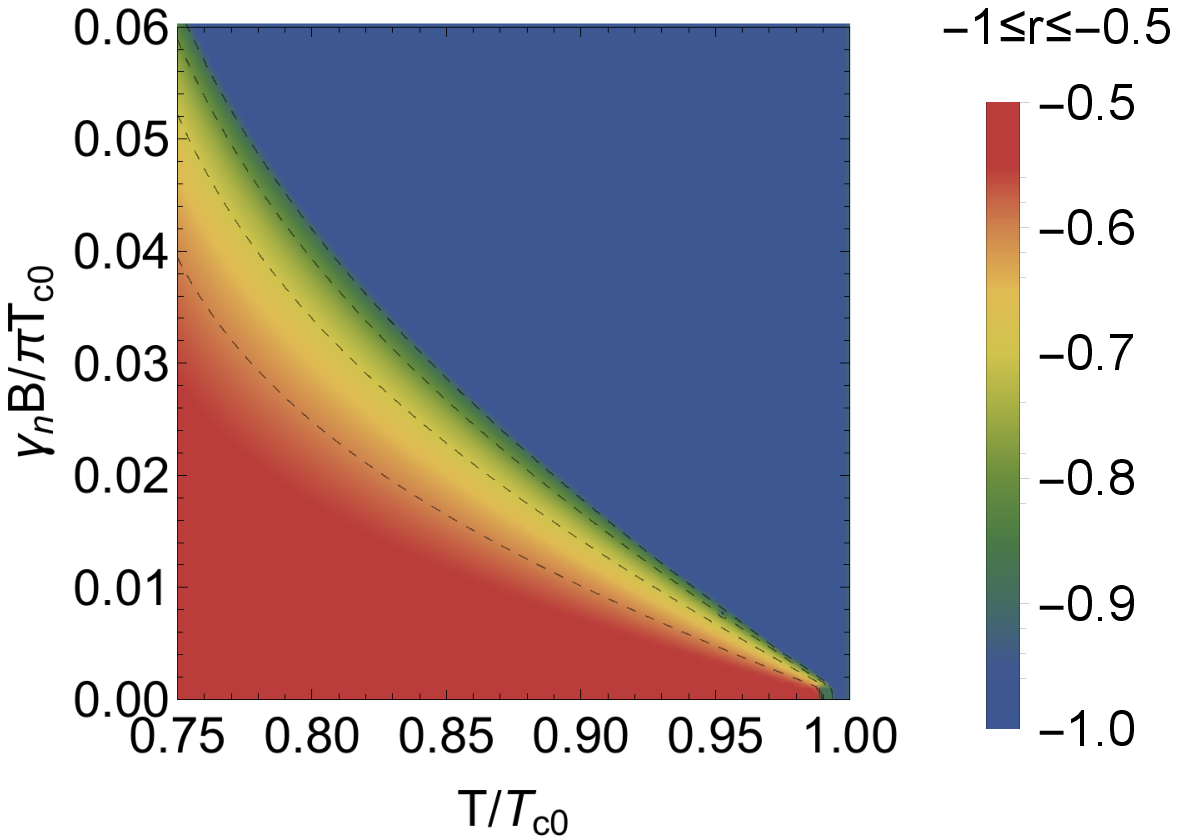}
\vspace{-1em}
\caption{The phase diagram on the $T$-$B$ plane. The left two panels (up-left and bottom-left) are the results up to ${\cal O}(B^{2}{A}^{2})$ in Ref.~\cite{Masuda:2015jka} (setting $\beta^{(4)}=\gamma^{(2)}=0$ in Eq.~(\ref{eq:eff_pot_coefficient02_f})), and the right two panels (up-right and bottom-right) are the results up to ${\cal O}(B^{4}{A}^{2})+{\cal O}(B^{2}{A}^{4})$ in the present study~\cite{Yasui:2018tcr}.}
\label{fig:r_B_T_MN_YCN_FL}
\end{center}
\end{figure}

For the nematic phase in the neutron $^{3}P_{2}$ superfluidity, we parametrize the condensate $A^{ab}$ as the diagonal form, $A(t,\vec{x})=A_{0} \, \mathrm{diag}\bigl(r,-1-r,1\bigr)$,
with two real numbers $A_{0}$ and $r$ whose ranges are restricted to $A_{0} \ge 0$ and $-1 \le r \le -1/2$~\cite{neutron_superfluidity_3P2}.
This restriction does not loose the generality.
We assume the uniform system, and hence suppose that $A_{0}$ and $r$ are independent of space and time.
Having this parametrization, we obtain several different phases: the UN phase for $r=-1/2$, the D$_{2}$-BN phase for $-1<r<-1/2$ and the D$_{4}$-BN phase for $r=-1$.
The phase realized in the ground state is determined by minimization of the free energy (\ref{eq:eff_pot_coefficient02_f}).
We show the phase diagrams on the $T$-$B$ plane by the temperature ($T$) and the magnetic field ($B$) in Fig.~\ref{fig:r_B_T_MN_YCN_FL}.
The magnetic field is switched on along the $y$-axis: $\vec{B}=(0,B,0)$.
We notice that the magnetic field in the magnetar is around $\gamma_{n}B/(\pi T_{c0}) \simeq 0.02$.
Thus the magnetar can cover the UN, D$_{2}$-BN and D$_{4}$-BN phases.
We compare our result with the previous one in which only the leading term of the magnetic field was considered ($\beta^{(4)}=\gamma^{(2)}=0$ in Eq.~(\ref{eq:eff_pot_coefficient02_f}))~\cite{Masuda:2015jka}.
We find that the region of the D$_{2}$-BN phase is extended by the higher order terms of $\beta^{(4)}$ and $\gamma^{(2)}$.
Notice that the order parameter in the D$_{4}$-BN phase is not affected by the magnetic field.
This should be expected reasonably, because the neutron pairing in the D$_{4}$-BN phase is the superposition state of the spin $\uparrow\uparrow$ pairing and the spin $\downarrow\downarrow$ pairing with equal fractions, and thus the energy shifts by the magnetic field should be canceled in the quasi-classical approximation.

Form Eq.~(\ref{eq:eff_pot_coefficient02_f}), we obtain the thermodynamic quantities, the heat capacity $C(T,B)$ and the spin susceptibility $\chi_{i}(T,B)$ for the spatial direction $i=1,2,3$~\cite{Yasui:2018tcr}.
They show the discontinuities at the phase boundaries, indicating the second order phase transitions, which is consistent with the result from the BdG equation~\cite{Mizushima:2016fbn}.
It is interesting that $\chi_{i}(T,B)$'s exhibit anisotropies due to the symmetries of the UN, D$_{2}$-BN, and D$_{4}$-BN phases, such as $\chi_{1}(T,B)=\chi_{2}(T,B)\neq\chi_{3}(T,B)$ in the UN phase and $\chi_{1}(T,B)\neq\chi_{2}(T,B)=\chi_{3}(T,B)$ in the D$_{4}$-BN phase.
In D$_{2}$-BN phase, all $\chi_{i}(T,B)$'s are different each other.
Such property may be useful to study the internal structure of neutron stars.

\section{Conclusion and Discussions}

We have discussed the phase of neutron $^{3}P_{2}$ superfluidity under the strong magnetic field.
Starting from the $LS$ potential between two neutrons, we have derived the Ginzburg-Landau equation around the transition temperature.
In the present study, we have calculated the higher order terms of the magnetic field which were not considered so far.
We have investigated the nematic phases on the $T$-$B$ plane, and found that the higher order terms of the magnetic field extends the D$_{2}$-BN phase.
We have calculated also the thermodynamic quantities, i.e. the heat capacity and the spin susceptibility.
Those information will be useful to research the interiors of neutron stars.
As future studies, it will be interesting to ask how the neutron $^{3}P_{2}$ superfluidity is faced with the other phases, such as a neutron $^{1}S_{0}$-wave superfluidity and a hyperon matter.
The connection to a quark matter is an interesting problem, because both phases share the topological properties, such as quantum vortices~\cite{Cipriani:2012hr,Kobayashi:2008pk}, gapless fermions~\cite{Mizushima:2016fbn,Mizushima:2017pma}, and so on. 
Applications to higher spin systems will be also interesting~\cite{Venderbos}.

This work is supported by the Ministry of Education, Culture, Sports, Science (MEXT)-Supported Program for the Strategic Research Foundation at Private Universities ``Topological Science" (Grant No. S1511006). 
C.~C. acknowledges support as an International Research Fellow of the Japan Society for the Promotion of Science (JSPS) (Grant No: 16F16322). 
This work is also supported in part by 
JSPS Grant-in-Aid for Scientific Research (KAKENHI Grant No. 16H03984 (M.~N.), No.~18H01217 (M.~N.), No.~17K05435 (S.~Y.)), and also by MEXT KAKENHI Grant-in-Aid for Scientific Research on Innovative Areas ``Topological Materials Science'' No.~15H05855 (M.~N.).

\end{document}